\documentclass[icarus,fleqn]{article}
\input epsf
\usepackage{amsmath}
\begin{document}
%%%%%%%%%%%%%%%%%%%%%%%%%
\title{Snapshot coronagraphy with an interferometer in space}
\author{Boccaletti A. (1,2), Riaud P. (2,4), Moutou C. (3), Labeyrie A. 
(2,4) \\
(1) DESPA, Observatoire de Paris Meudon, place J.Janssen, \\
92195 Meudon, France, Boccalet@despa.obspm.fr \\
(2) Coll{\`e}ge de France, 11 Pl. M. Berthelot F-75321 Paris, France \\
 Riaud@mesioq.obspm.fr\\
(3) ESO santiago, Chile, Cmoutou@eso.org\\
(4) Observatoire de Haute-Provence, F-04870 St Michel l'Observatoire, \\
France, Labeyrie@obs-hp.fr}
%\date{Received xxx; accepted yyy}
\date
\sloppy
\maketitle
\newpage
\begin{abstract}
Diluted arrays of many  optical apertures will be able to provide high-resolution snapshot images if the beams are combined according to the densified-pupil scheme. We show that the same principle can also provide coronagraphic images, for detecting faint sources near a bright unresolved one. Recent refinements of coronagraphic techniques, i.e. the use of a phase mask, active apodization and dark-speckle analysis, are also applicable for enhanced contrast. Implemented in the form of a proposed 50-500m Exo-Earth Discoverer array in space, the principle can serve to detect Earth-like exo-planets in the infra-red. It can also provide images of faint nebulosity near stars, active galactic nuclei and quasars. Calculations indicate that exo-planets are detectable amidst the zodiacal and exo-zodiacal emission faster than with a Bracewell array of equivalent area, a consequence of the spatial selectivity in the image. 
\end{abstract}
% *************************************************
\newpage
\section{Introduction}
\label{intro}
An early proposal for searching exo-planets involved coronagraphy aboard the "Large Space Telescope" (Bonneau {\it et al.} 1975) which became the Hubble Space Telescope (HST). It was shown that telescope rotation and synchronous image subtraction could remove much of the coronagraphic residue, a speckled halo of star light expected to outshine the planet image by 3 or 4 orders of magnitude. The corresponding Faint Object Camera (FOC) was built by the European Space Agency, but the COSTAR corrector subsequently installed aboard HST to correct the figuring error of its main mirror destroyed the size matching of the Lyot stop in the FOC coronagraph and made it inoperable. Also, the aperture size had been shrunk from 3m to 2.4m, and the telescope rotation proved difficult to achieve. The FOC could however have succeeded since exo-planets brighter than expected were recently found by radial velocity measurements (Mayor and Queloz 1995, Marcy and Butler 1996, Delfosse {\it et al.} 1998).\\ 

The rotation modulation principle was later adopted, together with a form of coronagraphic field-selective attenuation, by Bracewell and McPhie (1979) for their proposed interferometer exploiting the improved contrast of a planet in the  infrared. The concept has been refined by L\'eger {\it et al.} (1996) for the DARWIN proposal, and also adopted by Angel and Woolf (1997) for their Terrestrial Planet Finder. Both instruments are intended to observe in the 10$\mu$m infra-red. At this wavelength, the Earth's luminosity is quoted to be $7.10^{6}$ times, or 17 magnitudes, fainter than the Sun's. \\

Coronagraphic imaging is now also considered for the Next Generation Space Telescope (Gezari {\it et al.} 1997, Moutou {\it et al.} 1998, Trauger {\it et al.} 1998, Le F\`evre {\it et al.} 1998). The added refinement of "dark-speckle imaging" can further improve the detection sensitivity 10 to 1000 times (Boccaletti {\it et al.} 1998a), while the use of a phase mask and the active attenuation of residual starlight in the annular field push the sensitivity towards the exo-planet detection threshold at visible wavelengths.\\

In this article we analyze and confirm the suggestion (Labeyrie 1999b) that these techniques of apodized imaging are also usable with large multi-aperture interferometric arrays operating in the densified-pupil imaging mode. Indeed, it was recently shown that such arrays  operated in this mode can produce snapshot images in a narrow field (Labeyrie 1996). We have studied the coronagraphic schemes applicable to such imaging conditions and find them suitable for exo-planet observing at infra-red wavelengths. At visible wavelengths they can provide higher angular resolution than  NGST coronagraphy,  allowing observations of fast orbiting exo-planets.\\

Our numerical simulations of imaging with the "Exo-Earth Discoverer" (EED) sketched in Fig. \ref{eedschema} verify the expected imaging performance. The EED is a 36-element space interferometer with size $D$ of the order of 100m. Having free-flying telescopes (Labeyrie 1985, B\'ely {\it et al.} 1996), it can be considered as a possible precursor of much larger instruments  such as the  150 km Exo-Earth Imager proposed for the longer-term goal of making exo-planet portraits (Labeyrie 1999b). 
% *************************************************
\section{Science}
The proposed instrument is intended for imaging, in the infra-red, and possibly the visible, faint circumstellar environments with their diffuse component and their point sources, brown dwarf companions or exo-planets. Brown dwarfs are expected to be 10$^7$ to 10$^8$ times fainter in the optical range and 10$^{5}$ times in the near infra-red. Extra-solar planets are 10$^6$ times fainter than their parent star at 10 microns. Their detection may also be affected by the presence of extended emission, both zodiacal and exo-zodiacal in their vicinity.\\

Jets and disks, as well as extreme cases of circumstellar activity such as in SS 433 (Spencer 1979) are obviously also relevant to high-resolution coronagraphy. Extended sources in the vicinity of stars are of interest, especially at the birth and death stages of the star life: observations of nebulosities leading to the formation of a planetary system, ejected envelopes, accretion disks in binary systems, disks of small bodies and planetary debris will all benefit from high-resolution imaging coronagraphy, as well as high-velocity jets and supernova remnants.\\

High-angular resolution can also access the inner parts of active galactic nuclei, inside the dust torus, and coronagraphy can improve the rejection of the bright unresolved core. In objects such as M81, QSO's and possibly gamma-ray bursters, where a central black hole is suspected, both the transverse and radial components of the gas velocity may become observable with milli-arcsecond angular resolution and the associated field spectrography. Also interesting is the observation of rings around supernovae in nearby galaxies.
% *************************************************
\section{Concepts}
\label{concept}
\subsection{Pupil densification}
Fizeau interferometers, the equivalent of giant telescopes having a sparse mosaic mirror, produce an image, but its quality degrades catastrophically when the aperture size becomes much larger than the sub-apertures. A usable image can then be retrieved by densifying the exit pupil, i.e. distorting it to increase the relative size of the sub-pupils. Such instruments, which may be called "hyper-telescopes", evade a requirement long believed to be a "golden rule of imaging interferometry" (Traub and Davis 1982, Beckers 1997), namely that the exit pupil be identical to the entrance pupil. Instead, one preserves only the arrangement of sub-pupil centers, while magnifying each sub-pupil with respect to the inter-pupil spacings, thus making the exit pupil more densely packed than the highly diluted entrance pupil (Fig. \ref{pupil}).\\

The result is a combined image where the interference pattern is magnified, with respect to the broad diffraction peak contributed by the sub-apertures. This broad peak acts as a window which limits the field but concentrates the energy in the useful central part of the interference pattern. The pupil densification mixes two different scales in the image: the low-resolution scale of the sub-images, or images contributed by a single sub-aperture, and the high-resolution scale of the interference pattern, which retains the classical convolution behaviour on extended objects. When the array is phased on a point source, a peak appears in the interference function, thus providing a direct image within the window.\\
Depending on the baseline redundancy, snap-shot exposures obtained in this way can contain between $N$ and $N^2$ resolved elements, if $N$ is the number of apertures. Simulations of an exo-planet's image, showing clouds and continents, have been obtained assuming a 48 or 150-aperture "Exo-Earth Imager", a larger version of the EED with 150 km size (Labeyrie 1999b).\\

Here we concentrate on the shorter-term goal of obtaining unresolved images of  exo-planets, and the other goals mentioned in the previous section. It requires a smaller instrument, spanning 50m to 500m, so that the central source be unresolved, as required for phase-mask coronagraphy (Roddier and Roddier 1997, Guyon {\it et al.} 1999) and for dark-speckle imaging (Labeyrie 1995,  Boccaletti  {\it et al.} 1998a, Boccaletti {\it et al.} 1998b). \\
For a sparse array geometry allowing full pupil densification, we assumed an entrance aperture shaped like an "exploded" hexagonal paving (Fig. \ref{pupil}). The case of 36 elements, with concentric rings of 6, 12 and 18, was mostly used in the simulations, but a wider imaging field is obtainable with more elements, at no loss if the collecting area is conserved. Once densified, the 36-element  pupil resembles the aperture of the Keck telescope (see Fig. \ref{pupil}).\
 
% *************************************************
\subsection{Phase-mask coronagraphy with multi-aperture arrays}
Classical coronagraphs since Lyot (1939) have an occulting mask in the first focal plane of a telescope, and, in a relayed pupil plane, a "Lyot stop" diaphragm. Slightly smaller than the pupil, it removes the double edge ring caused by diffraction, which contains most of the light propagated from the Airy rings (Malbet 1996). The focal mask is opaque, and has to cover at least the 4 or 5 central rings of the Airy pattern for efficient nulling. What remains in the final re-imaged field is a few attenuated Airy rings, becoming broken into random speckles at increasing distances from the vanished Airy peak. The speckles result mostly from the residual bumpiness of the telescope's mirror, but obscurations of spiders,  segmentation, etc., on the pupil can contribute. \\
With Roddier and Roddier's proposal (1997) of a transparent phase-shifting mask, smaller than the Airy peak ($\approx 0.5{\lambda\over D}$), similar extinction is achievable, but the usable field extends closer inwards, down to the very edge of the Airy peak. The phase mask is basically narrow-band. It can however be multiplexed for a wide spectral coverage by using a polychromatic Bragg hologram as a reflective phase mask (Labeyrie 1999b).\\

When a star's Airy peak is properly focused and centered on a phase mask, most light is diffracted outside the geometric pupil, and  is thus removable by a  diaphragm, the Lyot stop, before re-imaging the focal plane. If made to match exactly the geometric pupil, including any central obscuration, spider arms and segmentation gaps, the Lyot stop can be nearly optimal. In the presence of obscurations or mirror segmentation such as shown on  top of Fig. \ref{pupil}, this is another advantage with respect to conventional Lyot coronagraphy, where the bright double fringe along edges requires masking a significant part of the geometric pupil area (typically 20\% of the diameter is occulted).\\ 
Phase-mask coronagraphy for single telescopes such as the NGST (Moutou {\it et al.} 1998) can be extended to the case of a densified-pupil interferometer. Whether the wavefront focused at the entrance of the coronagraph has natural continuity or is a dense mosaic of wavefront patches coming from widely spaced sub-apertures in the entrance aperture  makes no difference indeed. A non-resolved star is therefore "nulled" identically in either case.\\

Another new coronagraphic scheme, the Achromatic Interfero-Coronagraph (Gay J. and Rabbia 1996, Rabbia {\it et al.} 1998), also offers the capability of imaging a companion very close to the field center, and uses an ingenious solution for achromatism. Although not included in the present simulations, it would be of interest to assess its performance in the EED context. 
% *************************************************
\section{Numerical simulations}
\label{simu}
\subsection{The algorithm}
We have performed numerical simulations of the proposed interferometer and coronagraph, using various occulting and apodizing schemes. In accordance with the theory of densified-pupil interferometry (Labeyrie 1996), the complex amplitude distribution in the exit pupil is considered as a convolution of a sub-pupil with a "fakir board", i.e. an array of Dirac peaks representing the centers of the sub-pupils (the peaks can be complex numbers to represent phasing errors). The combined focal image of a point source is therefore a product of an interference function, the squared modulus of the fakir board's Fourier transform, and a diffraction function similarly transformed from the sub-aperture.\\ 

If the array is correctly phased, the interference function resembles the classical Airy pattern with its central peak, although the outer rings here become broken into speckles. The usual phasing algorithms serving for adaptive optics in monolithic telescopes, using wavefront slope or curvature measurements, are not applicable to long baselines, but other approaches using sharpness criteria have been found usable on Earth and in space (Pedretti and Labeyrie 1999, Pedretti 1999). \\ 
An extended incoherent source convolves the interference function, but affects negligibly the diffraction function if the interferometer is highly diluted. The  interference function  may therefore be considered as a spread function, while the  diffraction function behaves like an envelope or field-limiting window applied to the convolved pattern.\\

For simulating the  phase-mask coronagraph attachment, we introduced a $180^\circ$ phase shift in the central part of the star's Airy-like interference peak. The mask size is quite critical and  was adjusted by trial and error for minimal residual star light in the annular field. The Fourier transform of the resulting complex amplitude distribution is then multiplied by the Lyot stop, in most cases made to match exactly the geometric pupil, and again Fourier transformed to obtain the final image of the star. \\
Adding the planet's image requires a similar calculation to be made for the planet. In the plane of the phase mask, this involves the same interference function, but attenuated and shifted off-axis before multiplication with the un-shifted diffraction function and the phase mask function. Since the  planet's Airy peak falls  outside of the phase mask, the  mask affects very little the planet's final image. \\
Various sizes and shapes of the mask and pupil diaphragm were also experimented for optimizing the nulling of the star and its feet.
% *************************************************
\subsection{Comparison of various pupil configurations}
We have considered diluted arrays of 36 elements, having either circular sub-apertures or, better, hexagonal ones allowing a full densification in the exit pupil. Although the interference functions are identical in both cases, we find that a darker coronagraphic field is achievable with the fully densified pupil. We therefore adopted this optimal pupil configuration for simulating the coronagraph in the following. The pupils and corresponding images are shown in Fig.\ref{pupil}. \\ 

When working with hexagonal sub-apertures and full densification, the first-order secondary peaks of the star's interference function are exactly located on the envelope's  dark ring. The field area, expressed on the sky in units of the  array's Airy  area (squared Airy radius) is approximately equal to the number of apertures. The secondary peaks of the star's interference function have the same intensity as the central peak, but have first-order lateral chromatism. Any stellar companion or planet present in the main field or in the extended field also has side-peaks. In this respect, in monochromatric light, the central field always contains a replica of the planet's image if the central image is outside (see Fig. 5a).
% *************************************************
\subsection{Phase-mask size and shape}
Here the sub-apertures of the interferometer are assumed  perfectly phased. \\
The maximal nulling of starlight  by the phase  mask is reached when the amplitudes with opposite phases are balanced in the pupil, and this requires a careful adjustment of the phase mask diameter (Fig. \ref{maskh}, top). A circular phase mask appears better than an hexagonal one. The calculated intensity profile in the coronagraphic image of the unresolved star (Fig. \ref{maskh}, bottom) shows that the average attenuation of the annular field achieved by the phase-mask coronagraph is of the order of 20 to 50, a value comparable to that obtained by Roddier and Roddier (1997) for a monolithic telescope. Although not a large gain, this brings the average halo level to $\approx  2.10^{-4}$ relative to the peak of the unmasked star image, a welcome improvement before applying the further darkening steps discussed below.\\
Residual phase errors, both the intra-aperture errors caused by the bumpiness  of the mirror elements and the inter-aperture errors caused by imperfect adjustment of the optical path differences among them, have the same effect as in monolithic telescopes:  the feet of the diffraction pattern become distorted and intensified, both before and after the coronagraph. The intra-aperture bumpiness diffracts  mostly outside of the imaging field and therefore contributes little to the image degradation. In the 10 micron infra-red, the inter-aperture path differences can be kept small compared to the wavelength. Our simulations indicate that wavefront bumpiness contributes little to the background. If its scale size is 10 times smaller than a sub-aperture and its RMS amplitude  $59 nm$, corresponding  to $\lambda /8$  at $0.5 \mu m$ and $\lambda /170$ at $10 \mu m$), the bumpiness contribution to the $2.10^{-4}$ background is only $10^{-8}$. Subtracting a reference star image, obtained with a different bumpiness, therefore leaves a speckled background at $10^{-8}$ level (Fig. \ref{maskh}, bottom), and it can be smoothed if the actuators are re-ajusted many times during a long exposure. In the  dark-speckle mode, the analysis is performed differently. 
% *************************************************
\subsection{Effect of full or partial pupil densification}
Another critical parameter is the amount of densification in the exit pupil, and numerical simulations have again served to characterize its effect  (Fig. \ref{ecart}). We varied the sub-pupil size and calculated the nulling in the annular coronagraphic field, extending from the second dark ring of the masked Airy peak $\rho_m$, outward to the first dark ring of the sub-aperture's Airy peak $\rho_f$. With 36 elements, its area is 25 times larger than the area of the first dark ring $\rho_u$ in the unmasked Airy peak (Fig. \ref{maskh}, bottom). 
The residual intensity reaches  $2.10^{-4}$ with full densification. It increases 10 times for 50\% density. The optimal phase-mask size depends on the density and may be adjusted for optimal nulling\\
Narrow gaps between elements, amounting to a few percents of the element size, are difficult to avoid in practice, but will not degrade significantly the field darkness. This is another advantage with respect to the classical Lyot coronagraph, where such amplitude patterns on the pupil can be disastrous.
% *************************************************
\section{Application to exoplanet detection}
\subsection{Choice of wavelength }
The planet/star contrast improves a lot from the visible to the 10$\mu$m infra-red, owing to the drop of the star's Planck function and the peaking thermal emissivity of the planet. Also, residual wavefront bumpiness causes smaller phase shifts and less scattered light at longer wavelengths. However, the angular separation of an 8m  space telescope such as the NGST becomes sufficient to separate planets from their star in the visible, although not in all cases, and provides in principle the same planet detection performance as an EED of identical collecting area.\\ 
% *************************************************
\subsection{Performance of direct imaging}
Optimal data reduction would involve a combination of  the dark hole and dark speckle techniques mentioned below. The simulation routines are demanding of computer time however, and we have here simulated  long-exposure images, a less sensitive method which smoothes the boiling speckles and thus cannot exploit the nulling achieved in the dark speckles, but nevertheless demonstrates the detectability of planets. The results shown in Fig. \ref{result} assume 36  phased apertures of 0.6m, for a total collecting area similar to that of DARWIN (about 40 m$^2$) (L{\'e}ger {\it et al.} 1996, Mennesson and Mariotti 1997). We have simulated a twin of the solar system located at 20pc, with Venus, Earth and Mars  near a $m_v=6.33$ star. At $10\mu m$ the stellar magnitude is 11.5. The detector noise and photon noise were also included in the simulated 10 hours exposure. Planets $\approx 7.10^6$ times fainter than the parent star are directly detected after subtracting  a reference image.\\ 
Phasing errors, modelled to represent the bumpiness of the  mirror segments,  corresponding to $\lambda/170$ rms, are included  both for  the  planetary system and the reference star. A long exposure was fabricated by adding 60 images obtained with independant random phase maps, and this was repeated with different phase maps for the reference star.  With larger wavefront errors of $\lambda/80$ (at $10\mu m$), the scattered light level is about $10^{-7}$ and is therefore higher than the Airy peak of  Mars which remains undetected.\\
We have carried out the simulation at 3 different wavelengths (9.5, 10 and 10.5 $\mu m$). Secondary peaks feature a 1$^{st}$ order  chromatic dispersion and cannot be confused with the white primary peaks. Due to the interferometer geometry, an off-axis source has 3 peaks at $120^\circ$ (1 primary and 2 secondary).\\
% *************************************************
\subsection{Further gain with dark-hole and dark-speckle 
techniques}
Malbet {\it et al.} 1995, and Trauger {\it et al.} 1999, have shown how minor corrections of the wavefront shape, applied with active optics, can further improve the darkening of a coronagraphic image. Using a coronagraphic exposure, the dark-hole algorithm derives a phase correction map. Once applied to the actuators, it improves the average darkness of the selected annular field. \\  
The pattern of residual speckles  changes  randomly after each iteration, a consequence of noise in the servo loop. Although these speckles are here much fainter and slower than those caused by residual turbulence on Earth, the situation is similar, and dark-speckle imaging (Labeyrie 1995, Boccaletti {\it et al.} 1998a) can similarly improve the star-light rejection if thousands of  short exposures can be  exploited. The actuators can be re-adjusted every 10s for example, and  frozen in-between during the exposures.\\ 
Dark-speckle observations performed on ground-based single-aperture telescopes (Boccaletti {\it et al.} 1998b) with a CP20+ photon counting detector (Abe et al. 1998) have verified the theoretical expectations. Their extrapolation to the situation considered here suggests that exoplanets ($\Delta m(10\mu m)\approx 16$ to $19$, $\Delta m(1\mu m)\approx 22$ to $25$) can be detected with exposures lasting hours to a few tens of hours. A limitation is the imperfect cleaning of the image,  with a few stellar speckles survfiving near the center. These can be removed  with a reference star, but perhaps also by modifying the dark-hole algorithm.
% *************************************************
\subsection{Signal to noise ratio}
In addition to the residual pattern of star-light, the image provided by a densified-pupil interferometer, contains light from the zodiacal and exo-zodiacal clouds. The part of these extended sources which lies outside of the narrow imaging field, but within the resolution patch of the sub-apertures, contaminates the image through the feet of the interference function.\\

An exo-planet peak in the image, with $J_p$ photo-events  detected every second, is thus contaminated with : \\
1- $J_s$ events/s  from the residual speckled halo of starlight;\\
2- $J_{ez}$   events/s from the image of the exo-zodiacal cloud;\\
3- $J'_{ez}$ events/s  from the feet of the exo-zodiacal image (the cloud being larger than the field, there is a contamination of the field from the missing parts of the cloud, through the feet of the spread function);\\
4- $J_z$ events/s  from the zodiacal cloud;\\
5- $J_T$ events/s from the telescope thermal emission.
\smallskip\\

To calculate  these quantities, we consider   an entrance aperture  of size D with N sub-apertures of size d. It receives  $I_p$ photons/s per square meter from the planet; $I_s$ from the star;  $L_z$ from the  zodiacal  cloud, per steradian; and $L_{ez}$ per steradian from the exo-zodiacal  cloud of angular size $\phi$, assumed unresolved by a sub-aperture. \\
Since the planet's peak concentrates most entering photons, $J_p = I_p N d^2 $.\\
With the long-exposure mode using a reference star, the subtracted stellar contribution at the planet peak may be approximated by $J_s=I_s N d^2 g$.\\
The radial profile $g = \rho^{-\alpha}$ of starlight is estimated if $\rho$ is the axial distance and $\alpha$ the log slope of the coronagraphic halo, which is of the order of  $\alpha\approx 1.5 \sim 2$ according to the simulation results of Fig. \ref{maskh} (bottom). $\alpha$ decreases slowly as the number of speckle patterns is increased, as does the speckle noise.\\ 
The  remaining components may be expressed as :
\smallskip\\
$J_z = L_z  \lambda^2 $  in the zodiacal cloud ;\\
$J_{ez} = L_{ez}  \lambda^2  N d^2 D^{-2}$ in the image of the exo-zodiacal cloud; \\
$J'_{ez} = L_{ez} \phi^2  d^2$  in the halo of the exo-zodiacal image;\\
$J_T=L_T\lambda^2$ since the telescope thermal emission produces a background similar to the zodiacal cloud.\\

The last four quantities describe the respective contributions of the  central peak and the infinitely wide pattern of secondary peaks in the interference function. This pattern adds a halo to the image of the exo-zodiacal disc. A similar halo is also added to the image of the zodiacal continuum, and dominates it  completely.\\
These quantities are invariant when N varies at constant collecting area $ N d^2 $, except $J'_{ez}$ which decreases for increasing N. Increasing N,  which  is of interest to enlarge the field window, thus also improves the planet's discrimination against the halo of the exo-zodiacal image. In the $ N \rightarrow + \infty $ limit, the sparse array with densified exit pupil provides the same image as a filled telescope of similar diameter, although attenuated in the ratio of collecting areas.\\
In the long-exposure observing mode,  noise components are: 
\begin{itemize}
\item[1.]{ photon noise $P_n$ from the added photon contributions in the same speckle as the planet peak, amounting to $P_n= \sqrt{( 2J_s+ J_p+ J_z+ J_{ez} + J'_{ez}  )T }$ , where  $T$  is the total exposure  time (the factor 2  in the stellar term arises  from the subtraction of a reference image);}
\item[2.]{speckle noise $S_n$ from the stellar contribution, $S_n=\sqrt{2J_s^2 T^2 /n_s}$ (Angel 1994), if $n_s$ is the  number of speckle patterns added in the long exposure. For $n_s > 20$ the speckle noise becomes negligible;}
\item[3.]{read-out noise, $R_n= \sqrt{2\sigma_{ron} n}$, where $\sigma_{ron}$ is the detector's read-out noise per speckle area, whith n exposures made on both the object and reference star;}
\item[4.]{thermal emission within the telescope: Diner et al. (1991) calculated that a telescope at 70K provides, at $10\mu m$, a 10\% thermal contribution in the image of the zodiacal cloud. At 40K, the mirror temperature considered for DARWIN and TPF, the ratio is in the range $10^{-2}\sim 10^{-3}$. Here, this ratio is unaffected when the Fizeau images goes through the densifier optics, provided its temperature is not higher than the primary mirror's. The image contamination from thermal emission is therefore negligible for planet searching at $10\mu m$.}
\end{itemize}
The long-exposure signal/noise ratio is therefore :
\begin{equation}
S/N_{EED}={I_p N d^2 T \over \sqrt{P_n^2+S_n^2+R_n^2 }}
\end{equation}
or 
\begin{equation}
\begin{split}
S/N_{EED}=&I_pNd^2T[(2J_s+J_p+J_z+J_{ez}\\
&+J'_{ez})T+2J_s^2T^2/n_s+2\sigma_{ron}n]^{-1/2}
\end{split}
\end{equation}

The primary peak from the planet becomes attenuated if it moves towards the edge of the field. However this causes two secondary peaks to move inwards and become brighter. Because their location is known a priori, they can be added to the primary peak. The total signal thus obtained  is invariant with respect to planet position, the energy being conserved. And the added contributions from uniform sky background within the planet's peaks are similarly invariant for the same reason. The same applies to the added contributions from the star's residual halo if it has the same radial attenuation as the envelope. The chromatic elongation of the secondary peaks does not necessarily affect the addition of the planet peaks since it is removable with a field spectroscopy arrangement. Therefore, the $S/N$ is little affected by a planet's motion in the field.\\
  
%With respect to the angular distance, the primary peak is therefore attenuated by the envelope function. The total flux in the primary peak is thus a fraction $\gamma$ of $I_p$ which is equal to 1 for on-axis object and 0 when $\rho=rho_f$ (see Fig. 3 bottom). Owing to the shape of the array, 2 secondary peaks from off-axis objects appear very close to the center and contribute to the $SNR$ (see the case of Mars on Fig. 5). The total flux from the primary and secondary peaks is always equal to $I_p$.
In the dark-speckle mode, the stellar contribution to photon noise and speckle noise is decreased, as calculated by Boccaletti et al. (1998a). Also, no reference star is needed if the residual fixed speckles can be discriminated from planets by refining the dark-hole routine. Whether sensitive wave analyzer schemes, such as Zernike's phase contrast, can help remains to be investigated.
% *************************************************
\subsection{Comparison with Bracewell nulling}
In a multi-aperture Bracewell interferometer, light from a field extending over the resolution patch of the sub-apertures is transmitted to 
the single detector pixel, although with a field-dependant attenuation 
defined by the high-resolution transmission map of the beam-splitter 
arrangement. This map has a central dark minimum, designed to attenuate 
the star, surrounded by brighter speckles which can be 
rotated or "boiled" for generating a planet signal.   Like the procedure of Bracewell and Mac Phie (1979),
this is a single-pixel version of the rotation/subtraction algorithm originally proposed for exo-planet searching
with the Hubble Space Telescope by Bonneau {\it et al.} (1975) (see also Davies 1980). The angular transmission map is such that the average transmission $\tau$ of the  planet's light and zodiacal light  is of the order of  $\tau =0.2$.\\

With respect to the $J$ terms previously listed,  the zodiacal and exo-zodiacal contributions must now be multiplied by $N\tau$
since the detector pixel now receives all the energy from the part of these sources appearing in the sub-aperture's  resolved patch.
Also, the exo-zodiacal image vanishes, as well as the speckle noise.
Thus, the  signal/noise expression becomes:  
\begin{equation}
\begin{split}
S/N_{Bracewell}=&I_p N d^2  \tau T [( J_s  + J_p  \tau + J_z   N \tau \\
&+ J'_{ez} N \tau )T + \sigma_{ron} n]^{-1/2} 
\end{split}
\end{equation}
where $g$ now describes the  starlight rejection factor, expected to reach  $10^{-6}$ in DARWIN.

Increasing  the number of sub-apertures $N$ at constant collecting area 
$Nd^2$ changes only the zodiacal and exo-zodiacal  signals, which increase in proportion to $N$. Unlike the EED,
Bracewell interferometers should thus remain limited to few apertures.\\

Using these expressions, we compared the performance of an EED having 36 apertures of 0.6m, and a DARWIN of equivalent area, both located  at 1 A.U. from the Sun. The instrument's transmission combined with the quantum efficiency of the detector is 50\% at $10 \mu m$. An Earth at 20 parsecs provides  4 ph/$m^2$ in one hour of exposure.\\
At $ 10 \mu m $, the luminosity of the solar zodiacal cloud varies between $7,6.10^{-7} W/(m^2.sr.\mu m)$, at the ecliptic pole,
\sloppy
the value used by Mennesson et al. (1997) and $1,1.10^{-4}W/(m^2.sr.\mu m)$.
We used the median value ($7,6.10^{-6} W/(m^2.sr.\mu m)$ corresponding to 21 magnitudes per square arcsec.
The equivalent angular size $\phi$ of the exo-zodiacal cloud  is taken as $\phi=0.25$arcsec, corresponding to 5 A.U.
in the solar system, with $L_{ez}=10 L_z$. \\

A 30 hr observation in the narrow spectral band ($\lambda=10 \mu m, \Delta \lambda=0.5 \mu m$) provides a 0.5 $\sigma$  detection with DARWIN,
and  a 5.8 $\sigma$  detection with the  EED in the long-exposure mode with "boiling" speckles.
With an achromatic phase-mask allowing a wider $6-18 \mu m$ band, the DARWIN spectroscopy band,
the  signal/noise ratio is 50.8 for the  EED  and 4.7 for DARWIN.\\

The gain in the EED case results from two effects:\\
1- the fully constructive interference forming the planet peak in the image, as 
opposed to the 20\% average intensity in the transmission map of 
DARWIN.\\
2- the fact that the planet peak can be isolated  from  most of 
the zodiacal contamination in the high-resolution image.\\  

For a given collecting area, the array size does not affect the signal/noise ratio. With the EED, the image is magnified, within its fixed window,
when increasing the array size, and this can bring the planets optimally located in the  field.
Increasing the sub-aperture count N from 36 to 100, their size being kept constant,
increases EED's signal/noise ratio nearly  proportionally from 5.8 to 16.0.
Instead, DARWIN's value of 0.5 in 30 hours increases to 0.8, as the square root of the aperture number.\\    

% ************************************************

\section{Conclusion and future work}
The prospect for coronagraphic imaging with large interferometric 
arrays, at
infrared and visible wavelengths, opens vast areas of science. The 
simulations 
presented show that searches for exoplanets at 10$\mu$m  should be 
feasible once  the  formation flight of mirror elements in space becomes 
mastered. We recall the main conclusions :
\begin{itemize}
\item{ Pupil densification allows direct imaging and coronagraphic nulling.}
\item{ A phase-mask, replacing the classical Lyot opaque mask, 
cleans the image closer to the attenuated star, and is more tolerant to the unavoidable narrow gaps between sub-pupils
in the densified pupil.  Both achromatisation of the 
mask and low-resolution spectroscopy ($\lambda / \Delta \lambda = 20$) are desirable for analysing  the images obtained.}
\item{At given collecting area, more sub-apertures is better in terms of image field and signal/noise ratio. }
\item{The optimal wavelength is a trade-off between  resolution,  
stellar spill-off, planet contrast, background emission and detector 
performance. At visible wavelengths, a monolithic 8m telescope often 
resolves the planet from its star and then has the same planet-searching  
efficiency as an EED of identical area.}
\item{ In addition to planet searches, direct imaging with the EED is  also 
of interest for the other science goals briefly described above.}
\item{With respect to Bracewell's detection scheme, and its DARWIN and 
TPF proposed implementations,  the image formation in the EED 
improves markedly, at equal collecting area, the discrimination of  an exo-planet amidst the  zodiacal and 
exo-zodiacal nebulosity.  The EED however requires 36 apertures  at least,
but these are smaller and can be implemented with identical mirror segments compatible with the nano-satellite philosophy }.

\end{itemize} 
The coronagraphic image is compatible with field spectrography.

Additional simulations are needed to compare the possible exposure routines,
assess the desirability of a fine wavefront sensor, and refine the noise models.
More work is also needed to design optimal optical trains, including the
coronagraphic device and its spectro-imaging attachment.    
\\  

Acknowledgments : 
Jean-Marie Mariotti his constructive was among the first 
to understand our ideas and on densified-pupil imaging encouraged us to
explore their application.
His premature death deprived us of his insight
and critical appreciation, particularly regarding these results.  

\newpage

% *************************************************

\Large{\bf{Figure Captions}}\\[1cm]
\normalsize

\begin{figure}[t]
\centerline{\epsfxsize=7cm\epsfbox{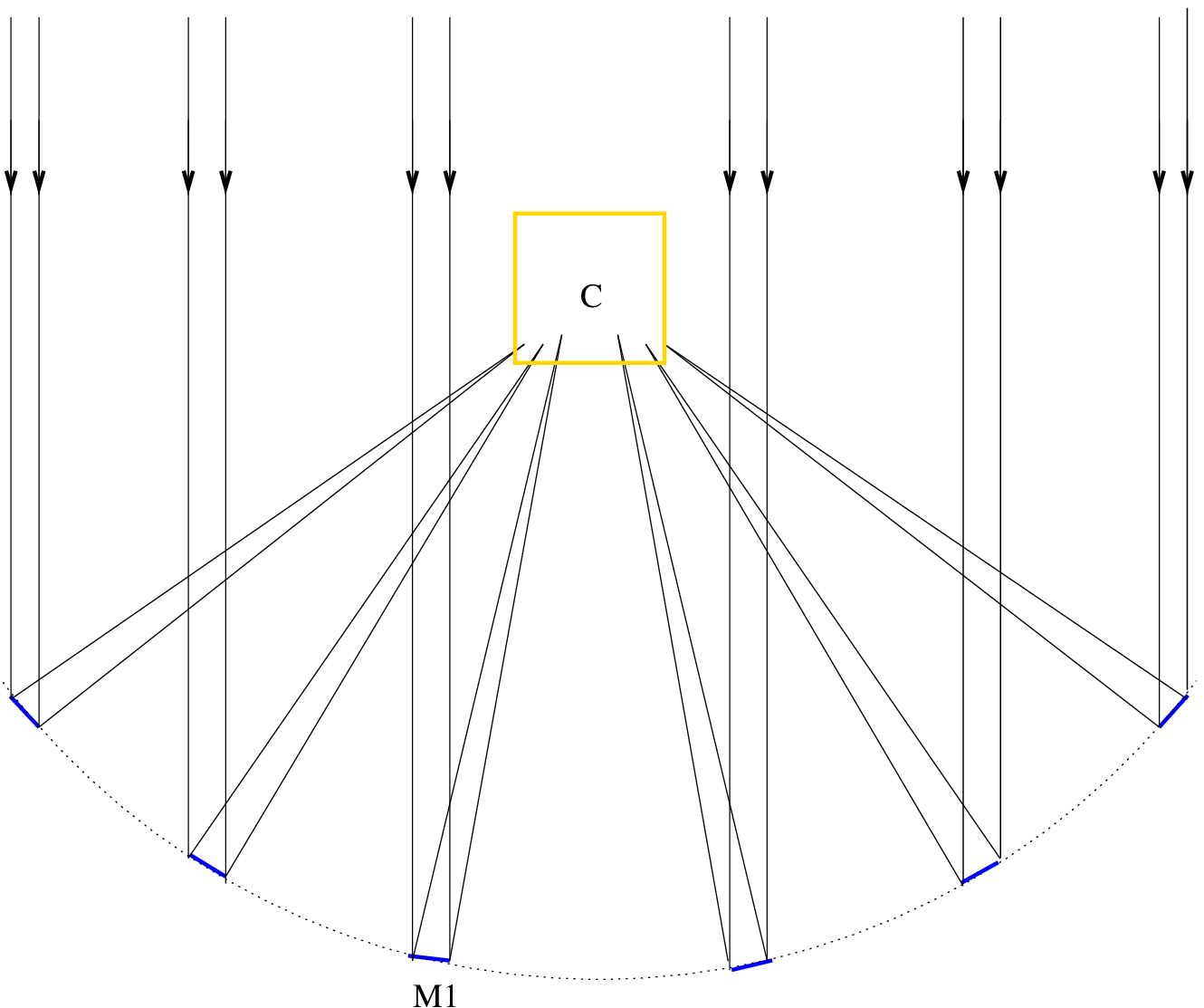}
\hspace{0.5cm}\epsfxsize=8cm\epsfbox{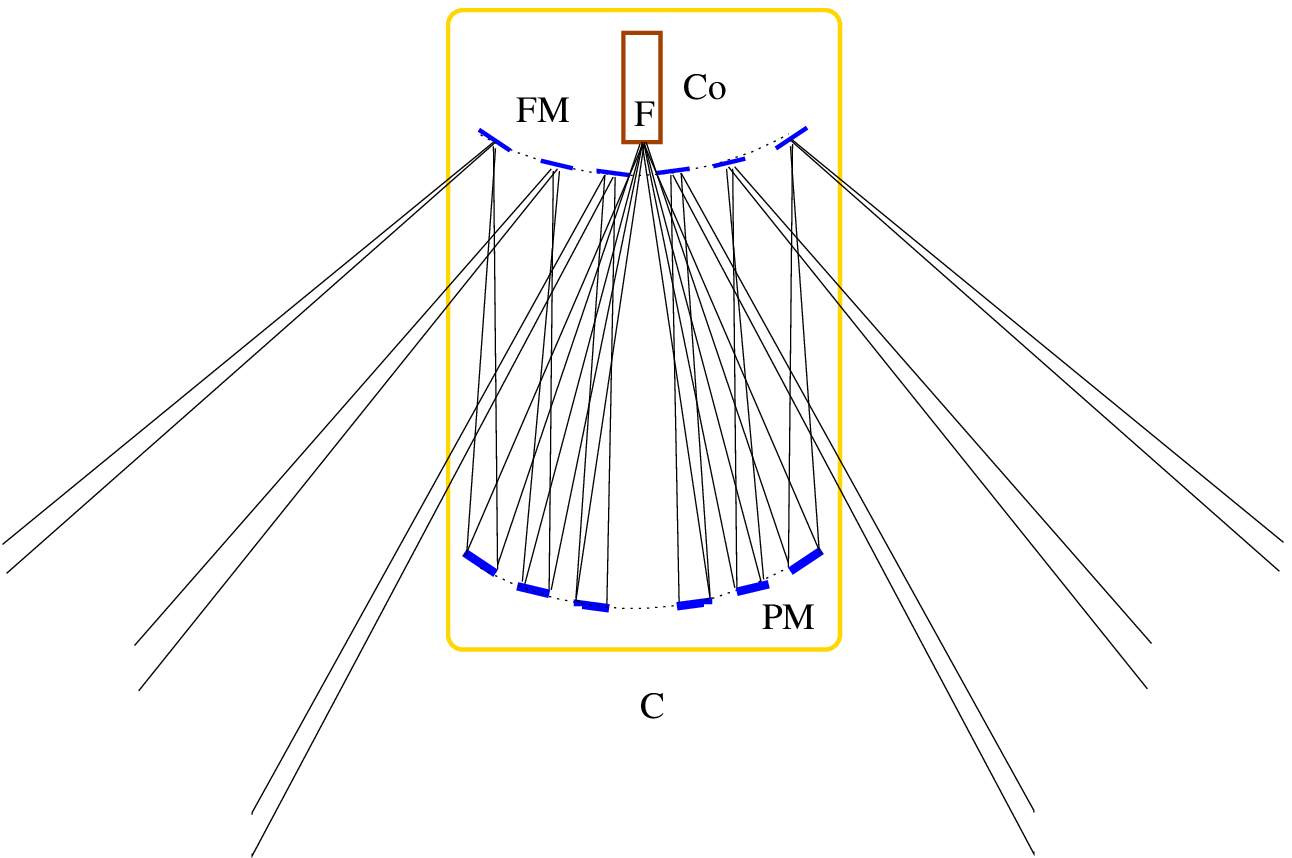}}
\caption[]{ }
\label{eedschema}
\end{figure}

Figure \ref{eedschema}: Possible optical train for the Exo-Earth Discoverer.
Mirror elements M1 focus starlight in the focal combiner C (detail below),
on field mirrrors FM, which form a densified image of the pupil on mirror segments PM.
A combined image is focused in F at the entrance of a coronagraph Co.
Mirrors PM are carried by actuators for tip-tilt and piston corrections.\\ 

\begin{figure}[b]
\centerline{\epsfxsize=6cm\epsfbox{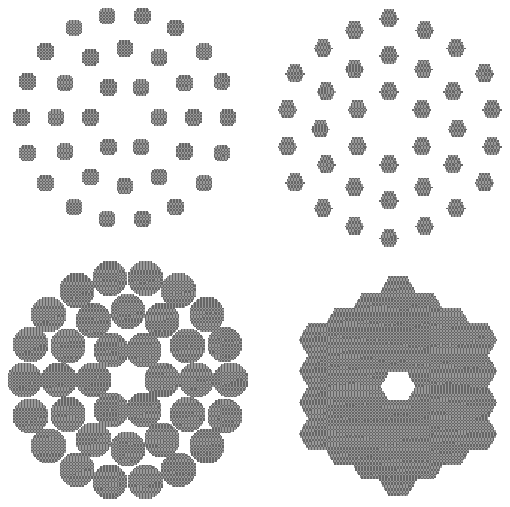}}
\centerline{\epsfxsize=7cm\epsfbox{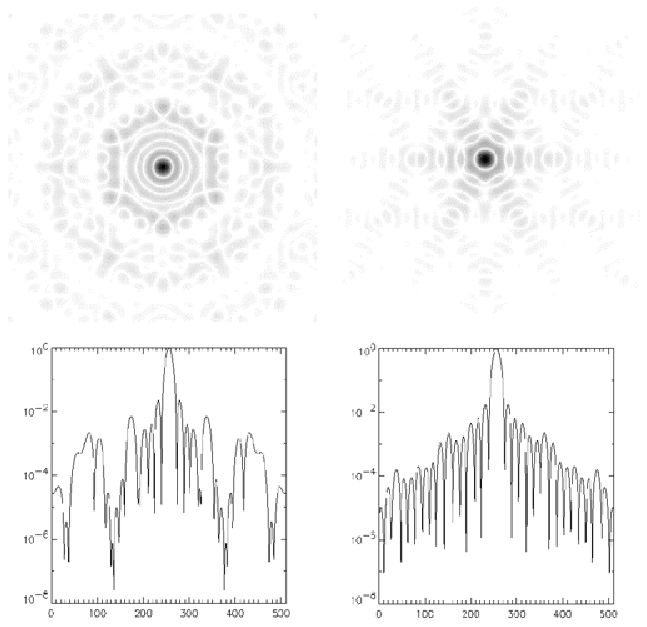}}
\caption[]{ }
\label{pupil}
\end{figure}

Figure \ref{pupil}: Interferometer apertures (top row)
with disk (left) or hexagon (right) subpupils, and densified version of them
obtained in the exit pupil (second row). The 100 to 200m Exo-Earth-Discoverer
has a sparse entrance aperture, but a fully densified exit pupil 
such as shown at right in the second row.
The star image (third row), with its diffractive envelope,
has most energy in the central Airy peak, as seen in the intensity profiles (bottom).\\

\begin{figure}[t]
\centerline{\epsfxsize=8cm\epsfbox{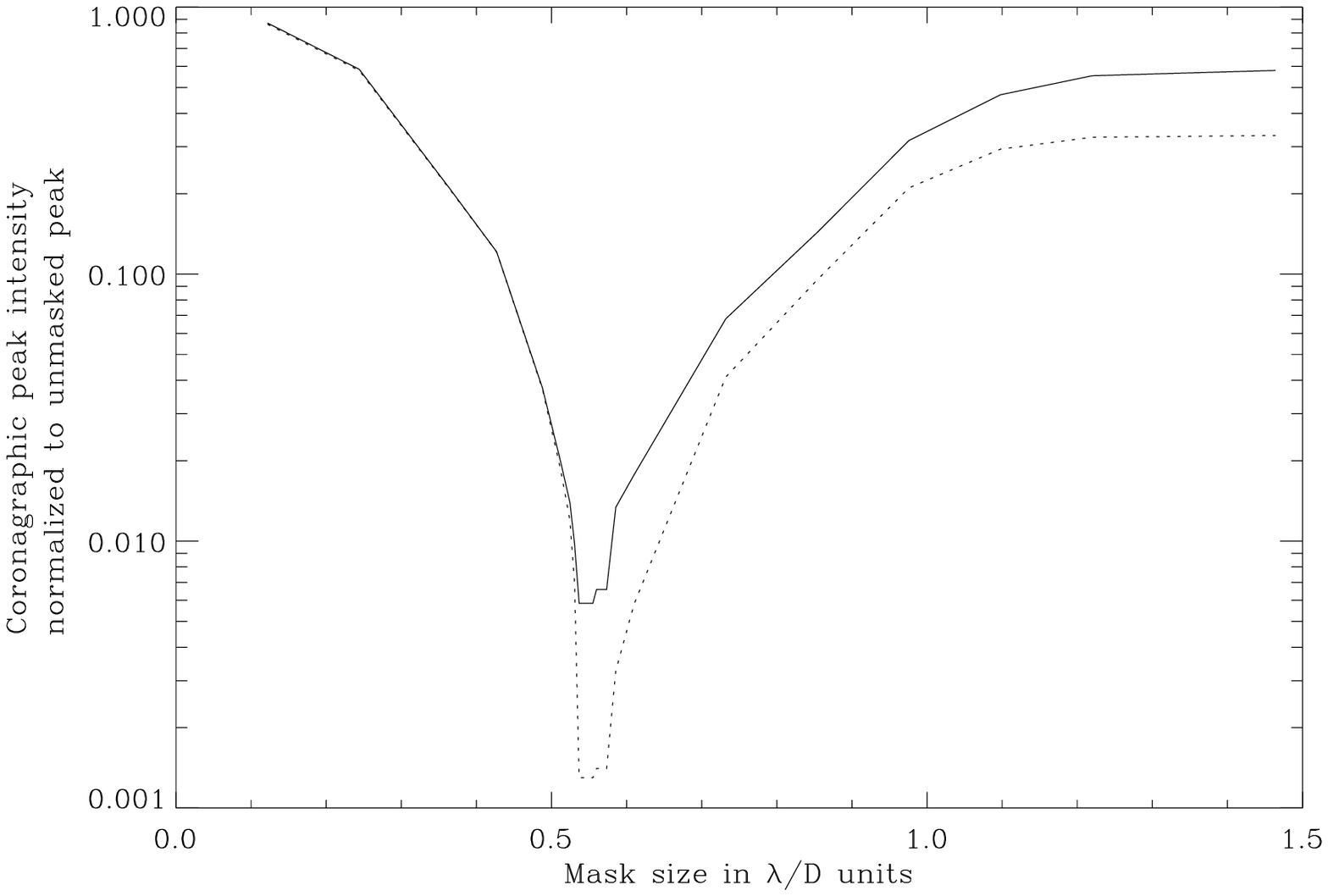}}
\centerline{\epsfxsize=8cm\epsfbox{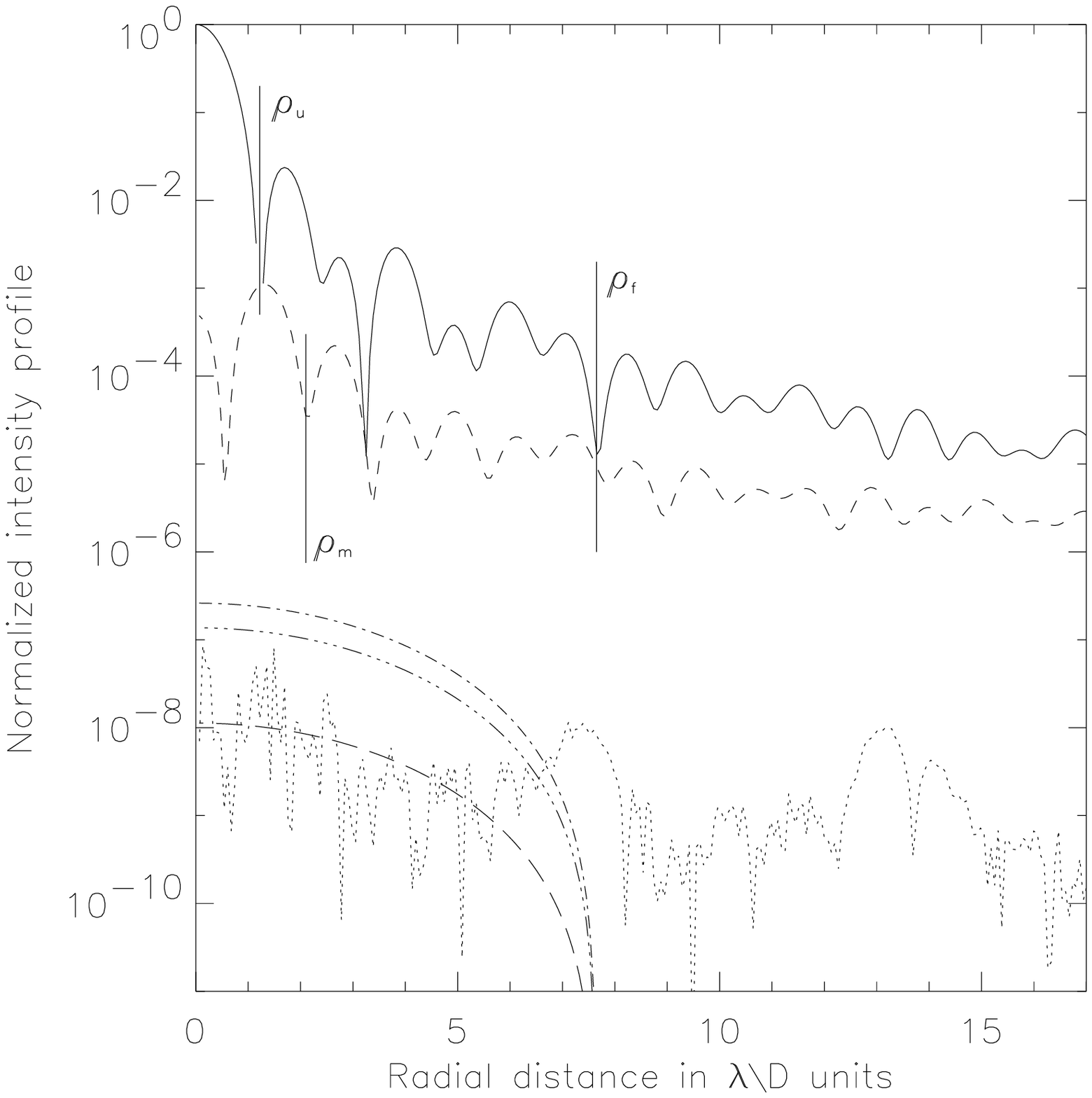}} 
\caption[]{}
\label{maskh}
\end{figure}

Figure \ref{maskh}: Top: Central darkening versus phase-mask size with a fully 
densified 
pupil of 36 adjacent hexagons, measured at the maximum ( dotted line) 
and across the area  (solid line) of the 
Airy peak . The  optimal mask size found is 
$(0.555 \pm 0.005)\lambda/D$. Bottom: Azimuthally averaged 
and normalized profiles of the unmasked Airy pattern (solid line), 
coronagraphic pattern (dashed line) and residual scattered pattern after 
subtraction of  the reference frame (dotted line), including photon and 
readout noise.  Both long exposures incorporated 20 different maps of
 ($\lambda/170$ rms) wavefront bumpiness. The radii of the 
first dark ring in the unmasked ($\rho_u$), the masked ($\rho_m$) and 
the sub-aperture ($\rho_f$) images are indicated. Also indicated is the 
level of Venus (dash-dot), Earth (dash-dot-dot-dot) and Mars (long 
dashes) multiplied by the sub-aperture's diffraction function .\\

\begin{figure}[t]
\centerline{\epsfxsize=8cm\epsfbox{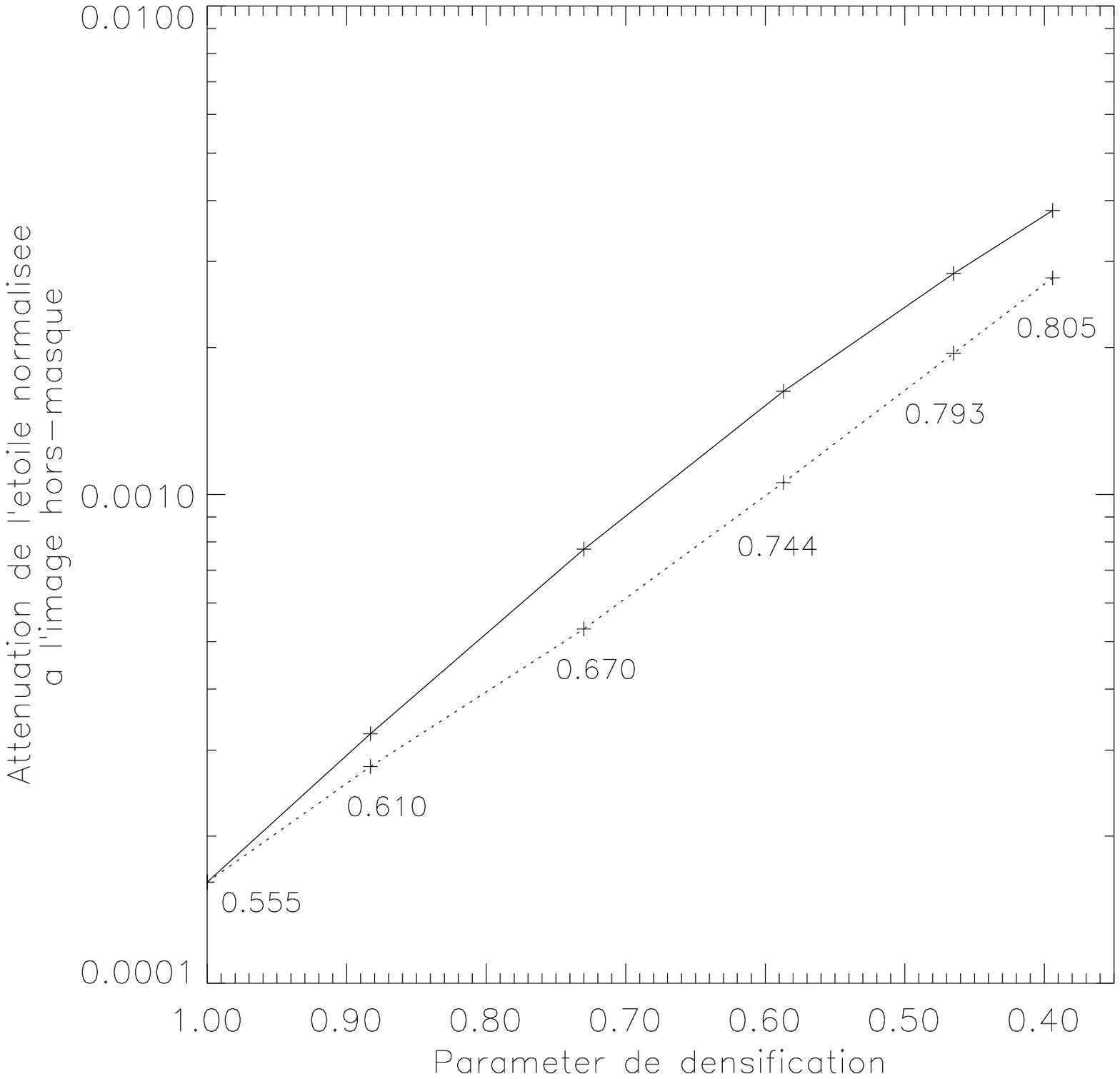}}
\caption[]{}
\label{ecart}
\end{figure}

Figure \ref{ecart}: Coronagraphic attenuation of star light versus exit-pupil 
density. The pupil density is measured as the relative area of the sub-
pupils in the exit pupil, while the attenuation is measured in the annular 
field ($12\lambda/D$) and normalized to the intensity of the unmasked peak. The dotted 
curve indicates that the mask size was re-optimized for each density 
value, while  the solid curve  indicates that the initial value 
($0.555\lambda/D$) was maintained . Numbers indicate the sizes of the 
re-optimized masks in units of $\lambda/D$.\\

\begin{figure}
\centerline{\epsfxsize=5.9cm\epsfbox{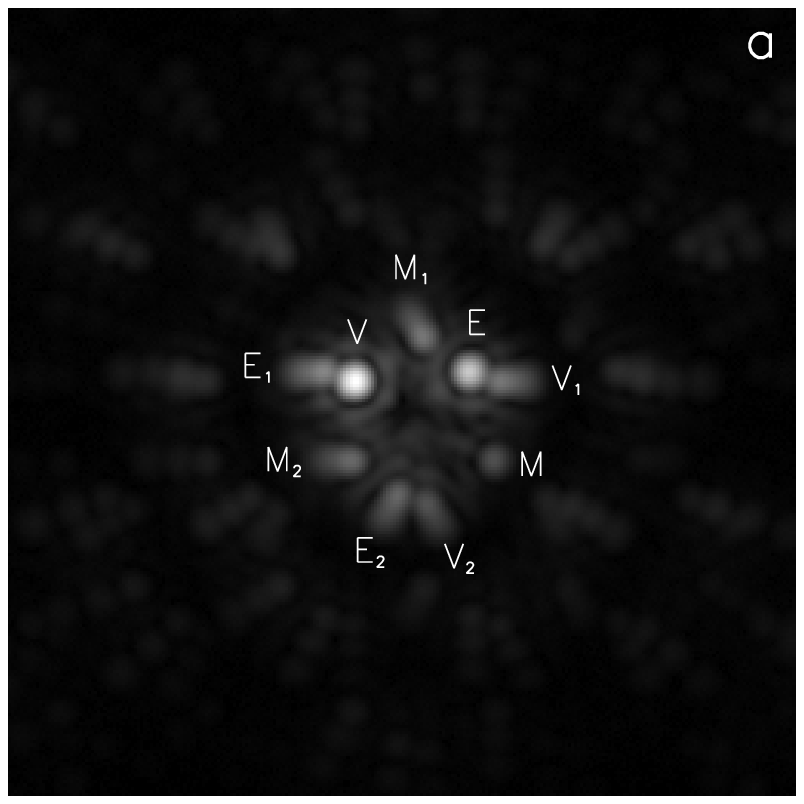}
\hspace{0.2cm} \epsfxsize=5.9cm\epsfbox{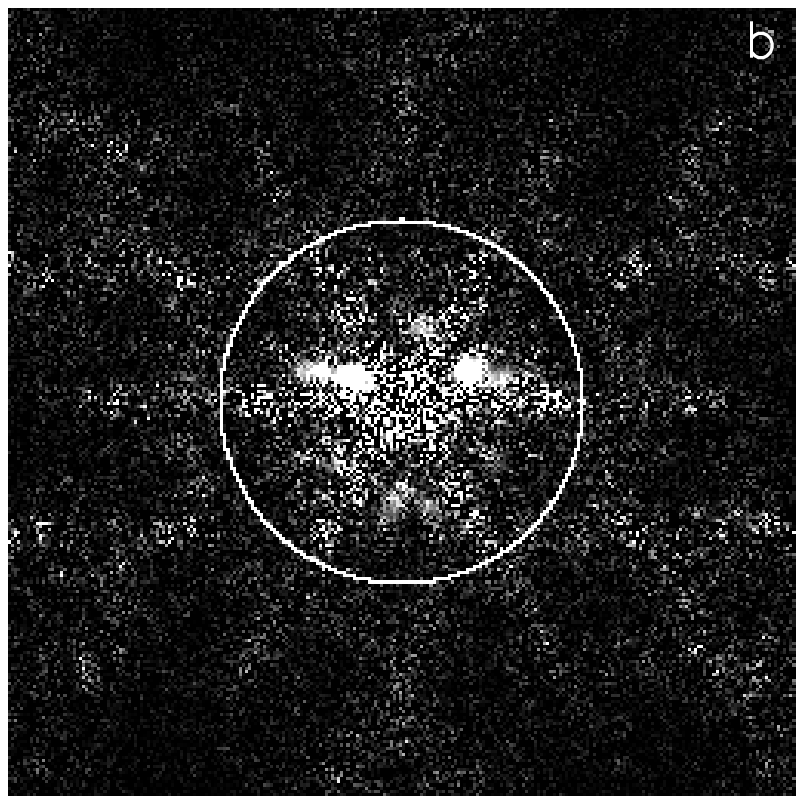}
\hspace{0.2cm}\epsfxsize=5.9cm\epsfbox{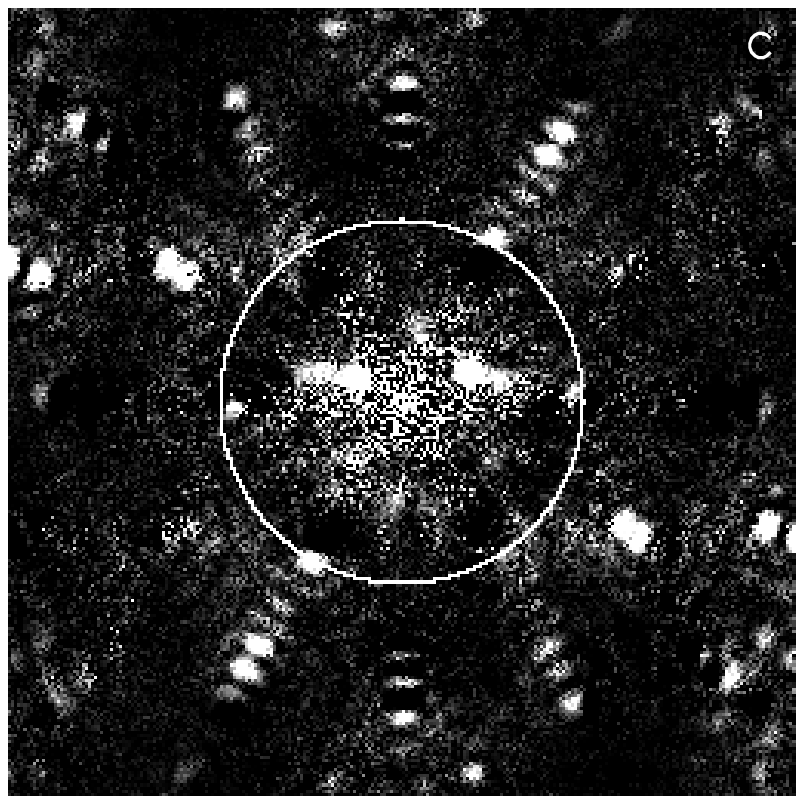}
}
\caption[]{}
\label{result}
\end{figure}

Figure \ref{result}: Coronagraphic image of a  solar system twin observed at 20pc ($m_v=6,33$) with the Sun attenuated to detect Venus, Earth and Mars.
The magnitude differences at 10 $\mu$m are respectively 16.4, 17.1, 19.8 relative to the parent star.
The simulated array has 36 telescopes of 0.6m, providing a collecting area similar to a DARWIN (40m$^2$).
To discriminate the dispersed secondary peaks from the white primary peaks, the final image was obtained
with a combination of 3 wavelengths (9.5, 10 and 10.5$\mu m$).\\
In (a), the star is removed to show the Venus, Earth and Mars components of the image, with their primary (V, E, M) and secondary ($V_1$, $V_2$, $E_1$, $E_2$, $M_1$, $M_2$) peaks.
The peripheral attenuation by the diffraction envelope makes the secondary peaks of Mars brighter than the primary peak, since they are closer to the field center.
No photon noise is present here.\\
The separations are $1.75\lambda/D$, $2.43\lambda/D$ and $3.69\lambda/D$, corresponding, for a 100m array, to 36, 50 and 76 milli-arcsecond at 10 $\mu$m, respectively.
A circle indicates the  field limit, at the first dark ring of the
diffraction envelope of a single 0.6m-aperture.
On the sky, the corresponding  field diameter is $12\lambda/D$, amounting to 0.24 arc-second at 10 $\mu$m with a 100m array of 36 elements.\\
Image (b) shows a 10 hours exposure with subtraction of a reference star image similarly obtained. Photon noise and readout noise (100 e-/pix/frame) were included.
A perfect wavefront was assumed. Image (c) is similar, but with 59 nm RMS wavefront bumpiness ($\lambda /170$), of 10mn lifetime. Two different sets of 60 phase maps served for the object and the reference star.
This level of bumpiness has little effect in the usable field, within the white circle.
A further gain in sensitivity is achievable in the dark-speckle mode.\\

% *************************************************

\onecolumn
%\newpage
\renewcommand{\topfraction}{1.}
%\newpage

\newpage

\newpage

\newpage

% *************************************************
\end{document}